# CP violation in $B_s^0$ mixing with LHCb

Georg Krocker on behalf of the LHCb collaboration
*Universität Heidelberg,*
*Physikalisches Institut,*
*Philosophenweg 12,*
*69120 Heidelberg,*
*Germany*

The determination of the CP-violating phase $\phi_s$ in $B_s^0 \to J/\psi\phi$ decays is one of the key goals of the LHCb experiment. Its value is predicted to be very small in the Standard Model but can be significantly enhanced in many models of new physics. An angular analysis of the decay $B_s^0 \to J/\psi\phi$ using about 340pb$^{-1}$ of data is presented resulting in a Standard Model compatible value of $\phi_s^{J/\psi\phi} = 0.13 \pm 0.18(\text{stat.}) \pm 0.07(\text{syst.})$ rad. An additional extraction of $\phi_s$ from the decay $B_s^0 \to J/\psi f_0(980)$ is performed for the first time in LHCb. A combination results in $\phi_s = 0.05 \pm 0.17(\text{stat.}) \pm 0.07(\text{syst.})$ rad. To resolve the fast mixing frequency is an essential prerequisite of this analysis. This is demonstrated by a precision measurement of $\Delta m_s = 17.725 \pm 0.041(\text{stat.}) \pm 0.026(\text{syst.})$ps$^{-1}$.

## 1. INTRODUCTION

The mass eigenstates of neutral mesons are not identical to the eigenstates of the weak interaction. The flavour eigenstates $B_s^0$ and $\overline{B}_s^0$ are superpositions of the two mass eigenstates, $B_L$ and $B_H$, which gives rise to the oscillation between the two states (Figure 1(a)). The frequency of this oscillation is given by the difference in the mass of the two mass eigenstates $\Delta m_s = m_H - m_L$. Likewise the decay width of the two mass eigenstates differs by $\Delta\Gamma_s = \Gamma_L - \Gamma_H$. In the Standard Model (SM) this difference is predicted to be sizable, $\Delta\Gamma_s = (0.087 \pm 0.021)$ps [1].

CP violation can theoretically arise in different places in the $B_s^0$ system, as CP violation in mixing, decay or in the interference of both. The $P \to VV$ decay $B_s^0 \to J/\psi\phi$ gives acess to the latter type. The CP violating phase $\phi_s$ arises due to interference of the tree level decay and the mixing via box diagrams shown in Figure 1. Higher order penguin contributions to this decay are expected to be negligible in the Standard Model. With this approximation $\Phi_s$ can be directly related to the CKM matrix elements

$$\phi_s^{J/\psi\phi} \approx -2arg\left(-\frac{V_{ts}V_{tb}^*}{V_{cs}V_{cb}^*}\right) \quad (1)$$

. Its value is very precisely predicted to be $\phi_s^{J/\psi\phi,SM} = (-0.0363 \pm 0.0017)$ rad [2]. In case of new physics this can be modified by an additional CP violating phase $\phi_s = \phi_s^{SM} + \phi_s^{NP}$.

The $B_s^0$ mixing frequency $\Delta m_s$ as well as $\Delta\Gamma_s$ and $\phi_s$ have been measured by the Tevatron experiments in the past but the precision is not accurate enough to rule out a deviation from the SM predictions [3][4]. LHCb, a dedicated $B$ physics experiment, can profit from the large number of $B$ mesons produced at the Large Hadron Collider to improve this results. The analysis steps and results of these measurements are discussed in the remainder of this article.

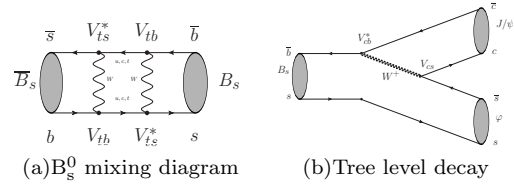

(a)$B_s^0$ mixing diagram  (b)Tree level decay

Figure 1: The decay of e.g. $B_s^0 \to J/\psi\phi$ can happen via the depicted tree level diagram (b) or via mixing (a) and subsequent decay. The Feynman diagram for $B_s^0 \to J/\psi f_0$ is similar to (b).

## 2. MEASUREMENT OF THE $B_s^0$ MIXING FREQUENCY

The measurement of the $B_s^0$ mixing frequency in the decay $B_s^0 \to D_s^-\pi^+$ is done in a blinded analysis [5] using an unbinned likelihood fit to the mass and proper time distribution of about 9200 signal $B_s^0$ candidates. The measured value is $\Delta m_s = 17.725 \pm 0.041(\text{stat.}) \pm 0.026(\text{syst.})$ps$^{-1}$. This measurement is compatible with the LHCb measurement done on 36fb$^{-1}$ of data taken in 2010 [6] and the measurement published by the CDF collaboration. The time dependent mixing asymmetry

$$A(t) = \frac{N_{\text{unmixed}}(t) - N_{\text{mixed}}(t)}{N_{\text{unmixed}}(t) + N_{\text{mixed}}(t)}. \quad (2)$$

of the $B_s^0$ signal candidates is shown in Figure 2.
As for all measurements that exploit mixing of neutral mesons a precise knowledge of the production and decay flavour is needed to decide wheter a given meson has mixed or not. In the used decay the charge of the final state particles is tagging the flavour of the $B_s^0$ at decay time. The flavour at production time is determined by dedicated flavour tagging algorithms which exploit the associated production of quark anti-quark pairs during the fragmentation of the b quark. The so called opposite side taggers (OST) attempt a partial reconstruction of the second B hadron in the event to define the flavour of the signal B. In case of the $B_s^0$ a same side kaon tagging (SST) algorithm is available additionally which uses the fact that an additional s quark is produced in the fragmentation of the $B_s^0$ that can form a charged kaon. The sensitivity of CP and mixing measurements directly depend on the effective tagging power $\epsilon_{\text{eff}} = \epsilon_{\text{tag}} D^2 = \epsilon_{\text{tag}} (1-2\omega)^2$ where $\epsilon_{\text{tag}}$ is the percentage of events for which the tagging algorithm was able to find a decision, $\omega$ is the percentage of wrongly tagged events and $D = (1-2\omega)$ defines the tagging dilution. The effective tagging power for the OST algorithms is measured to be $\epsilon_{\text{eff}} = 3.1\% \pm 0.8\%$ in the $B_s^0$ mixing analysis which is compatible to measurements in other B decays at LHCb. The efficiency of the so far not calibrated and not optimized SST is found to be $\epsilon_{\text{eff}} = 1.2\% \pm 0.4\%$.

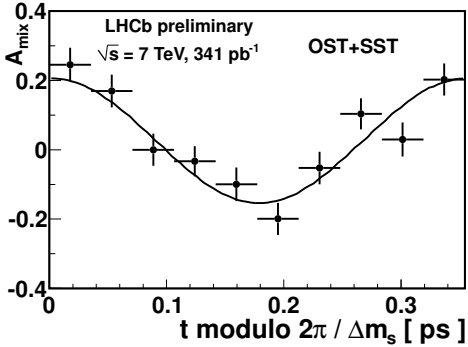

Figure 2: Mixing asymmetry for signal $B_s^0$ candidates as a function of decay time modulo $2\pi/\Delta m_s$. The fitted signal asymmetry is superimposed.

## 3. MEASUREMENT OF $\phi_s$ AND $\Delta\Gamma_s$

### 3.1. $\phi_s$ and $\Delta\Gamma_s$ in $B_s^0 \to J/\psi\phi$

The measurement of $\phi_s$ and $\Delta\Gamma_s$ in the decay $B_s^0 \to J/\psi\phi$ requires several inputs.

- The proper time resolution is measured by using $J/\psi$ originating directly from the interaction point and modelled with three gaussian functions with different widths but a common mean. The impact on the resolution of the $\phi_s$ measurement from this proper time resolution model corresponds to a single Gaussian with a width of $\langle\sigma_t\rangle = 50$fs as a resolution model. The analysis uses events passing an lifetime unbiased (86%) and lifetime biased (14%) selection. The proper time acceptance is determined from the relative efficiency of events passing the lifetime unbiased and lifetime biased selection.

- As $B_s^0 \to J/\psi\phi$ is a decay of a pseudo-scalar to two vector mesons the final state is not a CP eigenstate and a angular analysis is needed to disentangle the two CP states. A precise knowledge of the angular acceptances is thus essential for the measurement of $\phi_s$. The angular acceptances for the three transversity angles $\phi$, $\theta$ and $\psi$ are determined on simulated events and cross checked in the decay $B_d^0 \to J/\psi K^*$ which has a similar structure as the signal mode [7].

The measurement of $\phi_s$ and $\Delta\Gamma_s$ in $B_s^0 \to J/\psi\phi$ has been performed using approximately 8280 signal candidates. The calibrated oposite site taggers were used to determine the production flavour of the $B_s^0$. The signal events contain about 4% contribution from non resonant s-wave decays which are accounted for in the unbinned maximum likelihood fit performed to extract the physics parameters. The likelihood contours in the $\phi_s$ and $\Delta\Gamma_s$ plane are shown in Figure 3. The results from the fit are $\phi_s^{J/\psi\phi} = 0.13 \pm 0.18(\text{stat.}) \pm 0.07(\text{syst.})$ rad, $\Gamma_s = 0.656 \pm 0.009(\text{stat.}) \pm 0.008(\text{syst.})\text{ps}^{-1}$ and $\Delta\Gamma_s = 0.123 \pm 0.029(\text{stat.}) \pm 0.011(\text{syst.})\text{ps}^{-1}$ [8]. The dominating systematic uncertainty is caused by the angular acceptances and the background description. The results for $\phi_s$ and $\Delta\Gamma_s$ are in good agreement with the Standard Model prediction. With this measurement LHCb provides the most accurate measured values of $\phi_s$ and $\Gamma_s$ up to now as well as the first significant measurement of a non-zero $\Delta\Gamma_s$ value.

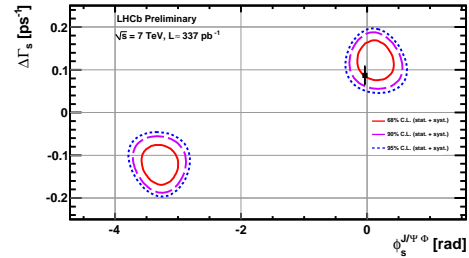

Figure 3: 2D likelihood confidence regions in the $\phi_s^{J/\psi\phi} - \Delta\Gamma_s$ plane. The black square corresponds to the theoretical predicted SM value.

### 3.2. $\phi_s$ and $\Delta\Gamma$ in $B_s^0 \to J/\psi f_0$

The decay of $B_s^0 \to J/\psi f_0(980)$ provides an additional mode for the measurement of CP violation in the $B_s^0$ system. The $s\bar{s}$ pair produced in the decay of the $B_s^0$ manifests itself in an additional resonance $f_0(980) \to \pi\pi$. This resonance has first been observed by LHCb and its relative branching ration has been measured to be $\mathcal{B}\left(B_s^0 \to J/\psi f_0\right)/\mathcal{B}\left(B_s^0 \to J/\psi \phi\right) = (21.7 \pm 1.1 \pm 0.7)$ [9]. The decay features a pure CP odd final state thus no angular analysis is necessary to measure CP violation. However, as a simultaneous measurement of $\Gamma$ and $\Delta\Gamma$ is not possible using only one CP state $\Gamma$ is used as input from $B_s^0 \to J/\psi\phi$. The first measurement of $\phi_s$ in this decay channel has been performed by LHCb using 1430 signal candidates. The likelihood contours in the $\phi_s$ and $\Delta\Gamma_s$ plane are shown in Figure 4. The result from the fit is $\phi_s^{J/\psi f_0} = -0.44 \pm 0.44(\text{stat.}) \pm 0.02(\text{syst.})\,\text{rad}$ [10]. A first naive combination of this result with $\phi_s^{J/\psi\phi}$ derived from $B_s^0 \to J/\psi\phi$ assuming Gaussian errors yields $\phi_s = 0.05 \pm 0.17(\text{stat.}) \pm 0.07(\text{syst.})\,\text{rad}$ [11]. Results of an analysis attempting to extract $\phi_s$ by using both decay modes in a simultaneous fit are expected soon.

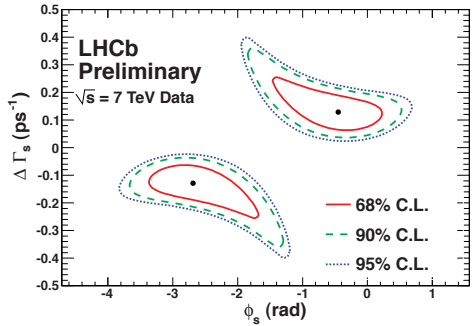

Figure 4: 2D likelihood confidence regions in the $\phi_s^{J/\psi f_0}$ plane. The black dots correspond to the best fit value.

## 4. SUMMARY

The measurement of the CP violating phase $\phi_s$ using 340pb$^{-1}$ of data collected with the LHCb experiment in 2011 has been presented. The tagged analysis in the decay channel $B_s^0 \to J/\psi\phi$ yields a value of $\phi_s^{J/\psi\phi} = 0.13 \pm 0.18(\text{stat.}) \pm 0.07(\text{syst.})\,\text{rad}$. The decay width difference in the $B_s^0$ system has been measured to be $\Delta\Gamma_s = 0.123 \pm 0.029(\text{stat.}) \pm 0.011(\text{syst.})\text{ps}^{-1}$. The phase $\phi_s$ was also extracted using the decay $B_s^0 \to J/\psi f_0(980)$ and the result is $\phi_s^{J/\psi f_0} = -0.44 \pm 0.44(\text{stat.}) \pm 0.02(\text{syst.})\,\text{rad}$. In addition the mixing frequency of neutral $B_s^0$ mesons has been determined to be $\Delta m_s = 17.725 \pm 0.041(\text{stat.}) \pm 0.026(\text{syst.})\text{ps}^{-1}$. All measurements are compatible with the Standard Model predictions and superseed revious measurements in accuracy.

## References

[1] Alexander Lenz, Ulrich Nierste, *Numerical updates of lifetimes and mixing parameters of B mesons*, Proceedings of the CKM workshop 2010 in Warwick, hep-ph/1102.4274v1

[2] CKMfitter Group (J. Charles *et al.*), Eur. Phys. J. C41, 1-131 (2005) [hep-ph/0406184], updated results and plots available at: http://ckmfitter.in2p3.fr

[3] CDF Collaboration (A. Abulencia *et al.*), *An Updated Measurement of the CP Violating Phase $\beta_s^{J/\psi\phi}$ in $B_s^0 \to J/\psi\phi$ Decays using 5.2 fb$^{-1}$ of Integrated Luminosity*, Public Note 10206

[4] CDF Collaboration (A. Abulencia *et al.*), *Observation of $B_s^0 - \overline{B}_s^0$ oscillations*, Phys. Rev. Lett. 97: 242003, 2006

[5] The LHCb Collaboration, *Measurement of $\Delta m_s$ in the decay $B_s^0 \to D_s^-(K^+K^-\pi^-)\pi^+$ using opposite-site and same-side tagging algorithms*, LHCb-CONF-2011-049

[6] The LHCb Collaboration, *Measurement of $\Delta m_s$ in the decay $B_s^0 \to D_s^-(K^+K^-\pi^-)(3)\pi$*, LHCb-CONF-2011-005

[7] The LHCb Collaboration, *Tagged time-dependent angular analysis of $B_s^0 \to J/\psi \phi$ decays with the 2010 LHCb data*, LHCb-CONF-2011-006

[8] The LHCb Collaboration, *Tagged time-dependent angular analysis of $B_s \to J/\psi\phi$ decays with 337 pb$^{-1}$ at LHCb*, LHCb-CONF-2011-049

[9] The LHCb Collaboration, *First observation of $B_s \to J/\psi f_0(980)$ decays*, Phys. Lett. B698: 115-122,2011

[10] The LHCb Collaboration, *Measurement of $\phi_s$ in $B_s \to J/\psi f_0(980)$*, LHCb-CONF-2011-051

[11] The LHCb Collaboration, *Combination of $\phi_s$ measurements from $B_s^0 \to J/\psi\phi$ and $B_s^0 \to J/\psi f_0(980)$*, LHCb-CONF-2011-056